\documentclass[prb,twocolumn,showpacs]{revtex4}
\usepackage{epsfig}
\usepackage{amssymb}
\usepackage{wasysym}

\begin{document}
\title{\bf Study of Percolative Transitions with First-Order Characteristics \\
 in the Context of CMR Manganites}
\author{J. Burgy and E. Dagotto}
\affiliation{National High Magnetic Field Lab and Department of Physics,
Florida State University, Tallahassee, FL 32306}
\author{M. Mayr}
\affiliation{Max Planck Institute f\"ur Festk\"orperforschung,
Heisenbergstra\ss e 1, 70569 Stuttgart, Deutschland}

\date{\today}

\begin{abstract}
The unusual magneto-transport properties of manganites are widely
believed to be caused by mixed-phase tendencies and concomitant
percolative processes. However, dramatic
deviations from ``standard'' percolation have been unveiled experimentally.
Here, a semi-phenomenological description of Mn oxides is
proposed based on coexisting clusters with $smooth$ surfaces, as suggested
by Monte Carlo simulations of realistic models for manganites, also
briefly discussed here. The present
approach produces fairly abrupt percolative transitions and even first-order
discontinuities, in agreement with experiments. These transitions may
describe the percolation that occurs after magnetic fields align the
randomly oriented ferromagnetic clusters believed to exist above the
Curie temperature in Mn oxides. 
In this respect, part of the manganite phenomenology
could belong to a new class of percolative processes triggered
by phase competition and correlations.
\end{abstract}
\pacs{ 71.10.-w, 75.10.-b, 75.30.Kz}
\maketitle

\section{Introduction}
The understanding of the self-organized behavior of transition-metal
oxides, such as manganites and cuprates, has developed into one of the
dominant scientific themes of condensed matter physics.
Phase coexistence and competition between different kinds of orders
involving charge, orbital, lattice, and spin degrees of freedom
leads to physical ``complexity'' as a characteristic of their behavior.
The balance between competing phases
is subtle, and small changes in composition, magnetic fields, or
temperatures  lead to large changes in the material properties.
In the context of colossal magnetoresistance (CMR),
these phenomena motivated the nanoscale phase separation
theory of manganites \cite{moreo}, and its
concomitant percolative transition from the metal to the insulator.
Many experiments have reported results compatible with phase
separation and percolation in Mn oxides \cite{uehara,faeth,review}.
Recently, CMR has been observed in simulations of simple models of
competing phases, between $T_{\rm C}$, 
the ordering temperature, and $T^*$$>$$T_{\rm C}$, the
temperature where coexisting clusters start developing upon cooling
\cite{burgy}. In the presence of magnetic fields the preformed 
ferromagnetic (FM) clusters align their magnetic moments, and they percolate.
This encouraging agreement between theory and experiments
suggests that the basic concept behind the CMR phenomenon --phase
competition-- has been unveiled.

\begin{figure}[t]
\epsfig{figure=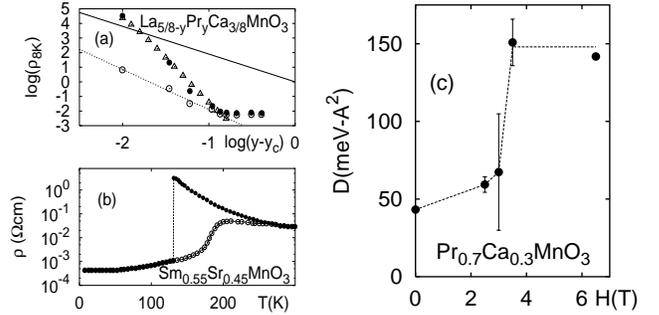,width=8.5cm}
\caption{ Results illustrating the ``abrupt'' character of the
metal-insulator transition in some manganites (compositions indicated).
(a) Dependence of residual $\rho_{dc}$ on chemical substitution at $0$
({\Large $\bullet$}) and 4~kOe ({\Large $\circ$}) \cite{uehara}.
The solid line represents the prediction of standard percolation, 
while the dotted one is a
guide to the eye.
The $\bigtriangleup$ are results of a simulation using a
broad distribution of bonds ($\alpha$=0.857).
(b) Resistivity vs. $T$,
at magnetic fields $0$ ({\Large $\bullet$})
and 7~T ({\Large $\circ$})  \cite{saitoh}.
(c) Field dependence of the spin-wave stiffness
at $T$=40~K \cite{baca}.}
\end{figure}

However, several experimental facts still await for a
rationalization within the phase separation scenario. Notorious among
them is the abrupt character of the metal-insulator transition (MIT)
in some manganites, result incompatible with {\it standard}
percolative phenomena where the transition is smooth.
The results in Fig.1 illustrate this paradox: 
(a) In $\rm La_{5/8-{\it y}} Pr_{\it y} Ca_{3/8} Mn O_3$, the
residual resistivity $\rho_{dc}$ at zero temperature $T$ 
changes fast with $y$, with a conductivity
exponent $t$ much larger than predicted by standard percolation
\cite{uehara}. 
(b) Far more challenging, 
$\rho_{dc}$ of $\rm Sm_{0.55} Sr_{0.45} Mn O_3$ 
changes from insulator
to metal, in a {\it first-order} transition \cite{saitoh}. 
Although most 
Mn oxides have continuous MITs, their $\rho_{dc}$ vs.
$T$ curves often show rapid changes and hysteresis loops,
in the same regime where cluster percolation seems to
occur. (c) Varying
magnetic fields, 
$\rm Pr_{0.7} Ca_{0.3} Mn O_3$ also presents a first-order-like MIT
\cite{baca}. These abrupt transitions cannot be rationalized
within standard percolation theory \cite{book}, but they are
found in regimes that exhibit some aspects of percolation \cite{review}.

The theory recently presented by Burgy et al. \cite{burgy} 
relies on a CMR state above $T_{\rm C}$ made out of preformed 
FM clusters, with random magnetic moment orientation (a similar
picture was also qualitatively discussed by Uehara et al. \cite{uehara}). 
This inhomogeneous state
is stable due to the presence of quenched disorder. The random
orientation of the cluster magnetic moments
can be achieved by having ``walls'' between
them made out of the competing insulating phase. Otherwise, when two
FM clusters with different orientations become in contact they
will align their moments. An insulating wall can prevent this process,
by drastically reducing the interaction between domains. However, 
when a relatively small
magnetic field is switched on, these preformed large FM moments
can easily rotate into the direction of that field. 
When this occurs the insulating walls are no longer needed to prevent
moment alignment, and they melt. This last process leads to a 
percolation, namely,
as the FM clusters rotate their moments they also slightly increase
their size to occupy the space left by the insulating walls,
allowing for charge conduction to occur\cite{burgy}. 
Although this last step appears percolative it is likely 
non-standard and abrupt, since it involves large rounded
clusters that are not very
ramified (contrary to standard percolation), due to surface tension
effects which are of relevance in a regime of phase competition.

It is the
main purpose of the present paper to describe, phenomenologically,
this percolation of smooth
objects, believed to describe CMR manganites.
It will be argued that first-order characteristics can be found in
percolative processes where surface tension effects are relevant.
It is important to clarify that the {\it full} description of the
CMR process is far more involved, with the random distribution and
subsequent field-induced alignment of
FM moments playing a key role. The first steps in that direction
were already described \cite{burgy}. Here our goal is the following:
if the FM clusters are assumed already aligned (and in that case the
spin degree of freedom can be dropped), how does the subsequent 
percolation occurs? It will be argued that this transition can be
fairly abrupt, even discontinuous.
The paper is organized as follows. In Section II
a variety of non-standard forms of percolation are described,
searching for situations where first-order transitions are induced. The main
conclusion of this section will be that achieving a discontinuous behavior
is highly nontrivial
and difficult. In Section III, a novel solution is proposed involving 
deterministic rules within a cellular automaton system. A discussion
of results is presented in Section IV, while Section V contains the
main conclusions.

\section{Forms of Percolation}
Standard percolation (SP) is a successful statistical model
for a variety of heterogeneous systems \cite{book}. 
It has been used to describe
diffusion through porous media, conduction
through random networks, and forest fires, among others.
In spite of its successes, however, SP
has several shortcomings, including the failure
to account for correlation effects. Indeed, in SP
each site (or bond) is occupied \emph{independently} of its
neighbors. This feature is responsible for the
ramified structure of the usual percolation clusters.
In Mn oxides
the coexistence of large clusters has been
reported \cite{uehara,faeth}, but the cluster distribution --while complex and
involving many lengths scales-- is not of the very ramified form
characteristic of SP. A possible explanation lies in
the long-range character of the elastic distortions induced by holes
in Mn oxides \cite{khomskii}, as well as surface tension effects as
described in the next section. 
These factors provide a natural tendency toward {\it smooth}
surfaces of coexisting metal-insulator clusters. In addition,
in physical systems with phase separation, there is
a local tendency to homogenize the landscape, minimizing
the interface tension.

\begin{figure}[t]
\epsfig{figure=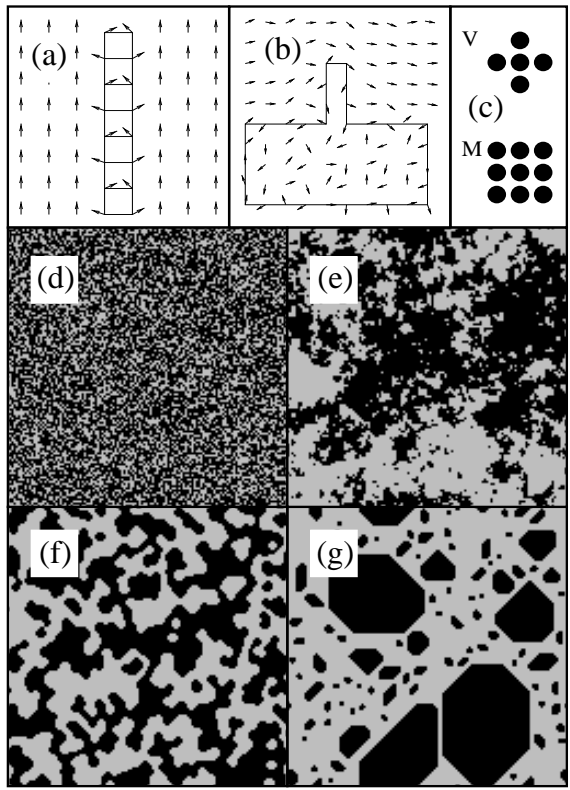,width=7.5cm}
\caption{ (a) Typical snapshot of an 8$\times$8  MC simulation of the
one-orbital model (Hund coupling $J_{\rm H}$=$\infty$, $T$=0.01$t$, $t$
the hopping amplitude,
and density 0.8). $J_{\rm AF}$ is 0 everywhere, but in the bonds of
the ladder where they take the value 0.12$t$. The average
spin-spin correlation in
the ladder is -0.4 (effective ``double-exchange'' hopping 0.54),
as opposed to -1.0 (effective hopping 0) when isolated from the
FM region (effective hopping 1).
(b) Similar to (a) but for density 0.5. Here $J_{\rm AF}$=0 is 0 in the
upper portion minus the finger-like region. In the rest $J_{\rm AF}$=0.15$t$,
favoring a flux state \cite{review}.
The effective hopping in the flux region is 0.7, while in the finger it
is as high as 0.9.
(c) The different neighborhoods in Vichniac's
notation \cite{vichniac}.
(d) Typical configuration for
standard percolation on a $128^2$ lattice at (or just below)
the critical density.
(e) Same as (d) but for correlated disorder ($\gamma$=0.2).
(f) Same as (d) but for the annealing rule (MGE5).
(g) Same as (d) but for the nucleating rule (MGE4).
The latter is particularly promising since a small increase in $p$
leads to a fully saturated final state, in a first-order
transition.}
\end{figure}

While simulations of models considering the elastic effects
through cooperative Jahn-Teller phonons are CPU time expensive,
the cluster rounding effects appear even in simpler Hamiltonians, such
as the one-orbital model \cite{review} (the actual form of this
Hamiltonian has been extensively discussed in previous literature, and
it will not be reproduced here). Monte Carlo (MC)
results in Fig.2(a,b) illustrate this phenomenon. In the simulations,
the Heisenberg coupling $J_{\rm AF}$ between localized $t_{2g}$ spins
is selected such that some regions of the studied lattice are in a
FM state if they were isolated, 
while others present some competing spin
arrangement (antiferromagnetic (AF) or flux \cite{review}). 
However, when the FM and non-FM regions
are in contact, there is a considerable influence from one another, and
the system tends to homogenize its transport properties. For example,
the ladder in Fig.2(a) would prefer to be in a perfectly AF spin state if
isolated.
However, by immersing the ladder into a ferromagnet, the spins cant
appreciably, allowing for substantial
charge hopping across the ladder. A thin AF region cannot
survive the ``pressure'' of a FM environment, and it transforms into
a fairly good conductor. Fig.2(b) shows similar evidence, 
now using competing FM and flux states. A finger-like region
emerging
from the lower-half, with coupling favoring the flux phase, also does
not survive in its original form but spins rotate to allow for
conduction \cite{comment}.
Modeling 
percolative manganites with 
the metal/insulating character of a link independent of its neighbors
can only be a crude approximation, and it
may cause the present discrepancy between
theory and experiment. If the clusters could
be made more `rounded' the transition would
be more abrupt, and hopefully in better agreement with real manganites.
The exponent $t$ may depend on the
short-scale real-space metal-insulator arrangement,
and changing the typical cluster shapes
may alter $t$ as well.



The physics of manganese oxides, in particular the abrupt change in
resistivity as a function of both field and temperature, 
is supposed to emerge from the realistic one- and two-orbital models
that are widely studied \cite{review}
(supplemented by the addition of quenched disorder).
%
%
%
However, simulations using realistic Hamiltonians for manganites
--containing the rounding effect spontaneously-- are simply
prohibitive in the percolative regime if critical exponents are needed, 
and one must rely on
semi-phenomenological approaches, as in the present paper. 
Even if we accept this simplification, there is, {\it a priori},
no obvious way to introduce the proposed local cluster compactification in
percolative scenarios. 
In alternatives to standard percolation such as
``bootstrap'' percolation \cite{bootstrap}, one simply
removes metallic sites or links if they do not have enough metallic
neighbors. This is repeated until no more sites can be removed.
However, numerical evaluation of the cluster number exponent $\tau$ 
suggests no new universal behavior \cite{chaves}. 
To generate compact configurations, 
long-range ``correlated disorder''
was also investigated here \cite{stanley}. 
The disorder is selected to follow the real-space correlation
function $C(\ell)$=$(1+\ell^2)^{-\gamma/2}$,
which falls off like $\ell^{-\gamma}$ for large distances $\ell$ but is
well-behaved at $\ell$=$0$. 
Typical configurations
confirm the presence of large clusters (Fig.2(e)). In addition,
our conductivity
studies for $\gamma$=0.2 (Fig.3) are compatible with a new
universality class and $t$$>$1.31=$t_{\rm SP}$, as predicted before
\cite{comment3}. However, to reproduce the large exponent $t$$\sim$7 
found in experiments \cite{uehara} anomalously small values of $\gamma$
are needed. The reason is that even with correlated disorder 
the weakest links still govern the critical  behavior
(Fig.2(e) shows weak links separating large clusters).
Correlated disorder, with its statistical nature that allows for those links,
is not sufficient to induce the smooth clusters 
needed to describe manganites.
Naively, these already known results appear to rule out the
possibility of finding an abrupt percolative transition to mimic
the behavior of manganites. 


Another alternative is the Swiss-cheese model \cite{cheese}. 
Halperin {\it et al.} \cite{cheese} have shown that this
model is equivalent to a random resistor network with a broad
distribution of conductances $g$ given by 
\begin{equation}
P(g) = (1-p)\delta(g)+p(1-\alpha)g^{-\alpha}.
\end{equation}
Here $P$ denotes the probability that a conductance has
 values between $g$ and $g+\delta g$, $p$ is the metallic
fraction as before, and $P(g)$ is
characterized by an additional parameter $\alpha$. 
The weakest bond 
dominates in the effective resistance of the backbone.
Extensive numerical tests \cite{octavio} support the conjecture
that $t$=$1/(1-\alpha)$, for 0.25$\le$$\alpha$$<$1.
The previous equation becomes an equality for  $\alpha$ large enough.
Hence, since the data in Fig.1(b) follows a straight
line with slope $-7$, it can be fitted with results of a
simulation at $\alpha\approx 0.857$. Although this numerical success is
encouraging, it is still intuitively unsatisfactory and does not address
the first-order percolative transitions of Figs.1(b,c). 
In fact, the
rapid change in $\rho_{dc}$ in Fig.1(a) may also be indicative of a 
first-order transition hidden by imperfections in the samples used.




\section{First-Order Percolation}
The above examples clearly show that it is highly nontrivial to generalize
percolation to yield a first-order transition, although it is possible
to change the critical exponents.
The existence of experiments (Figs.1(b,c)) with discontinuous 
transitions indicate that there must exist yet 
another class of percolative transitions escaping our attention. 
We believe that the key issue missing in standard percolation and the
related variations described above is a mimic of the effect of {\it
surface tension}, which naturally arises when two phases compete.
This naturally penalizes the roughness of a surface and avoids the
weak links ubiquitous in continuous percolation.
%
%

The simplest way to model the gregarious
tendency of the two competing phases of manganites
is to consider the nature of the cells
in the neighborhood of the cell under inspection.
If there are enough ``friendly faces'' around, the cell will not
change, otherwise it will flip to the competing phase.
In a magnetic system, the neighborhood induces a local field that
influences the center spin. In addition,
including oxygen movement, it is impossible to have an 
undistorted octahedron in the middle of a distorted JT region.
The phenomenological approach
followed here is based on Vichniac's majority voting cellular
automata \cite{vichniac}. Loosely speaking, cellular automata
are discrete dynamical systems whose evolution is described by
a local deterministic rule: the fate of every
cell at time $t$+1 is determined by a function of the configuration
of a few of its neighbors at time $t$. 
Cellular automata describe an approach to
equilibrium in a system dominated by strong, short-range
interactions.
\begin{figure}[t]
\epsfig{figure=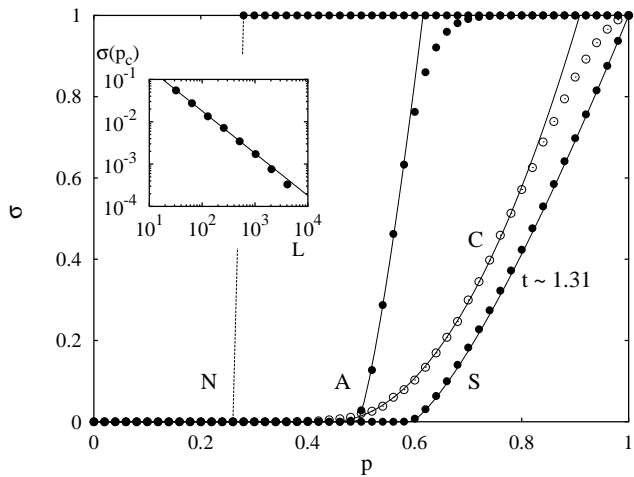,width=8.5cm}
\caption{Conductivity of a 2D resistor network vs. initial metallic
fraction, for several representative cases. N, A, C, and S stand for
nucleated (MGE4), annealed (MGE5), correlated, and standard, respectively. 
Both the A and S lines are proportional
to $(p-p_c)^t$, with $t\sim 1.31$. The inset shows finite-size scaling
data for A to support universality. $L$ is the linear size. 
Line C has a correlation power
$\gamma = 0.2$ and exponent $t$$\sim$2.1.}
\end{figure}
Vichniac \cite{vichniac} defines four different neighborhoods: Q
contains
only the nearest neighbors, V consists of Q plus the center
cell, H consists of Q plus the second-nearest neighbors,
and M consists of H plus the center cell (see Fig.2(c)). Unlike
in bootstrap percolation \cite{bootstrap}, the center cell is
treated on the same footing as its neighbors in V and M. This turns out
to be a \emph{key} property of the percolative process discussed here, 
making clusters compact and avoiding weak links.
This is quite different from bootstrap percolation
where the center cell is singled out.
After the neighborhoods are defined, the update rules are as follows. 
Given a site $i$ at a time $t$, let $S$ be the
number of occupied sites in the neighborhood $N$ of $i$. $S$ can be
any number between $0$ and the number of sites in the neighborhood.
If $S$ is greater than or equal to some threshold $T$, site $i$ will
be occupied at time $t + 1$. Otherwise it will be empty. The procedure
stops when no more changes occur \cite{comment2}. 
A rule is specified by a unique $N$ and $T$. 
This is referred to as NGET \cite{vichniac}. 
For the four neighborhoods
defined above, only $8$ rules (with $T$ close to half of the number of
sites in $N$) are nontrivial, as explained below.

The starting point for the calculations are always standard percolation
configurations, characterized by an occupation probability $p$. At time
$t$=0, $p$ is a good measure of the occupied fraction, within
statistical fluctuations. Under the ``time evolution'' provided by the
cellular automaton, the occupied fraction needs no longer be equal to
$p$. Rules with large (small) $T$  tend to (increase) decrease it.
For initial states with $p < p_c$, the final state does not
percolate. Two examples, to be discussed in more detail in the next
paragraph, are showed in Figs.2(f,g). Note the clear visual
difference from the more canonical cases Figs.2(d,e). The clusters
are now more rounded, and our intuitive goal of phenomenologically
mimicking the effect of surface tension has been reached.
After the time evolution stops, the resistivity is
measured using a numerically exact algorithm
\cite{franklobb}.

The increase in conductivity 
$\sigma$ immediately after $p_c$ ($= 0.5$) is clearly
faster in the annealed (MGE5, i.e. each site remains or becomes metallic if
five or more of its nine neighbors --including itself-- are metallic) case 
than in standard percolation, as Fig.~3
indicates. This is in agreement with our intuition that rounding effects
would speed up percolation, mimicking better the experimental
situation after the randomly oriented FM clusters align in a magnetic field.
However, it is surprising that this does not affect the value of $t$,
as the inset shows. 
Nevertheless, this rapid increase could be experimentally relevant. 
On the other hand, the \emph{nucleating rule} (MGE4, i.e. each site remains
or becomes metallic if four or more of its nine neighbors --including itself--
are metallic) leads to different results.
Its resistivity jumps from $0$ below $p_c$, where there is no connection
across the system (see Fig.2(g)), to the fully saturated value
above $p_c$. To understand this, note that the clusters in Fig.2(g)
have very clean boundaries. Adding a single metallic site immediately
outside one of the clusters will flip all the sites touched by the
boundary.
At $p$ close to $p_c$, it is likely that this apparently small
perturbation will
cause the cluster to coalesce with a neighboring cluster. In practice,
this starts a chain reaction (avalanche) 
that only stops when the entire lattice turns metallic.
{\it The nucleating procedure indeed provides a percolative transition 
that is first-order.}

How common are the first-order characteristics expected to be
in the cellular automata models described here?
Using the four (Q, V, H, M) neighborhoods defined above on the square
lattice,
there are eight non-trivial rules (QGE2, QGE3, VGE3, HGE4, HGE5, MGE4,
MGE5, MGE6).
By non-trivial we mean rules whose transition occur at a finite value of
$p$ (i.e. neither at $0$ nor at $1$) in the infinite system limit. In
all cases a careful size extrapolation was carried out numerically to
determine the critical concentration. They are respectively ($0.16$,
$0.83$, $0.5$, $0.33$, $0.66$, $0.25$, $0.5$, $0.75$). Although these
values are close to the fractions ($1/6$, $5/6$, $1/2$, $1/3$, $2/3$,
$1/4$, $1/2$, $3/4$), we know of no
rigorous result that they are equal to them. In two cases not listed
above (VGE2 and
VGE4) the numerical uncertainty prevented us from determining for sure
whether the rule is trivial or not. To be specific, the critical $p$ for
VGE2 is very small, and it becomes smaller still as we increase the
size of the
system used in the simulation. We could not rule out the fact that $p$
may extrapolate to zero in this case. As a consequence, we chose to
leave those two cases out of the present discussion. Of the eight
rules, the conductivity of three of them (QGE2, HGE4 and MGE4) was
found to vary discontinuously at the transition. The conductivity of
MGE4 is shown in Fig. 3 where the reader can convince oneself that
the transition is discontinuous. QGE2 and HGE4 look identical apart for
a shift in the location of $p_c$. Phenomenologically, the three cases
behave similarly in the sense that states very close to the transition are
unstable against flipping even just one site. Hence, we conclude
that first-order tendencies are certainly not a fragile feature among 
the set of possible voting cellular automata.
However, far more work is needed to fully understand the behavior of
these novel systems, with the present paper hopefully providing the first
steps in that direction. The detailed analysis of the models introduced here
may be as rewarding as the many previous studies of standard percolation.

\section{Discussion}

The novel discontinuous percolative transition described here can be applied
phenomenologically to the Mn oxides. In a previous related effort,
Mayr et al. \cite{mayr} constructed random resistor networks using
standard percolation and provided a tentative phenomenological description of
manganites. The curves $\rho_{dc}$ vs. $T$ mimicked properly the results
corresponding to some
manganites, but were too smooth to accommodate the materials of Fig.1 as
special cases. The nucleating-rule procedure to generate metal
and insulator clusters described here may solve this problem. The
left panel of Fig.~4 contains the effective resistivity (obtained
solving the Kirchoff equations) vs. $T$ for configurations generated by
the nucleating-rule, where the light (dark) regions in Fig.2(g) 
are insulating
(metallic). The individual resistivity of those regions is taken from
experiments, and they carry a $T$ dependence,
similarly as in Mayr {\it et al.} \cite{mayr}. The new
curves are in {\it excellent} agreement with experiments (see, e.g.,
Fig.1(b)), including the discontinuity,
and the fact that there is 
a common value of $\rho_{dc}$ at both low- and high-$T$.

\begin{figure}[t]
\epsfig{figure=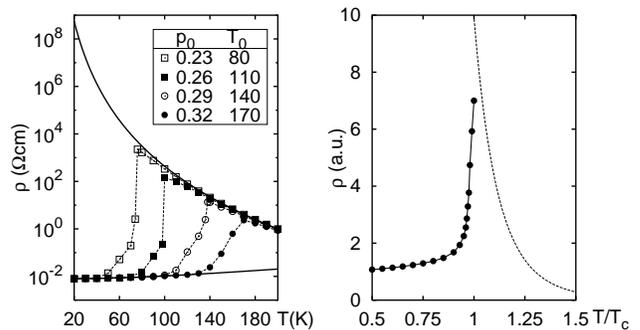,width=8.5cm}
\caption{
\emph{Left panel}: Resistivity $\rho_{dc}$ vs. $T$ using the nucleating
rule. The values of $\rho_{dc}$ for metallic and insulating bonds
are taken from experiments \cite{uehara}, as in \cite{mayr}.
Following \cite{mayr}, the metallic fraction varies linearly with
temperature, $p-p_0 = \alpha(T_0 -T)$. We consider $\alpha$=1$\permil$,
$T_0$=$100$K, and label the curves by $p_0$, their
metallic fraction at $T$=$T_0$.
\emph{Right panel}: $\rho_{dc}$ vs. $T$ 
of the $Q$=$20$ Potts model (for details, see text).
A cluster algorithm \cite{cluster} was used on a 
$64^2$ lattice. Long simulations (several thousand sweeps) were
required to sample both peaks in the probability distribution
function satisfactorily. Above $T_c$ there is no percolation
($\rho_{dc}$=$\infty$), but this can be solved phenomenologically by
allowing for conductivity among the several $Q$ clusters, 
as in the dashed line.
}
\end{figure}

Note that the new family of percolative processes presented here
naively seems triggered only by the occupation fraction $p$, not by
temperature, as opposed to some of the results of Fig.~1. However, 
the ferromagnetic fraction in real manganites is a function of 
temperature: clusters are expected to be formed at $T^*$ (although they
are not sufficiently large in size to create a global FM state), their number
grows with decreasing temperature (i.e. $p$ increases in the
language of this paper), until the Curie temperature is reached. It 
would be incorrect to believe that our results are only valid for
cases where the temperature is held fixed while concentrations are changed.

Another clarification is important. Why do all manganites not have
either a first- or second-order transition? Current experimental 
evidence suggests that both cases can be found, although the latter
tends to be fairly abrupt \cite{review}. Several aspects may be 
influencing these results. First, it may occur that available 
single crystals are affected by extrinsic effects in their
preparation. From this perspective, the current 
experimental results
may vary in the future as cleaner samples are prepared. 
But there are other issues to
consider. Clearly, different manganites have different degrees of
intrinsic disorder, such as the chemical disorder 
emerging from having different
ionic sizes at the A-site of the perovskite. This disorder will likely 
influence the clusters shape, and both configurations Fig.2(f) 
and 2(g) may emerge at different compositions, leading to either rapid but
continuous or first-order percolative processes, respectively.
It is clear that the issue of universality of the first-order transition
in manganites is far from clear both experimentally and theoretically
and more work should be devoted to its analysis. Here, we have simply
provided a plausible rationalization of some steps involved in the CMR
transition of the materials mentioned in Fig.~1. Whether all CMR
materials have the same kind of first-order transitions or not  will
hopefully be clarified by future experiments.

Finally, although the discussion here is based on a cellular automaton,
we found it reassuring that first-order percolative processes
can also arise from well-defined Hamiltonians. In particular, 
consider the $Q$-states Potts model known to have a a first-order
transition for $Q$$>$4 \cite{baxter}. Simulating the model 
($Q$=20) with MC techniques and using the Fortuin-Kasteleyn
mapping \cite{fortuin} to construct clusters for the various $Q$'s,
their percolative properties were analyzed \cite{fortunato}. Clusters for
each $Q$ are individually assumed metallic, and the  network 
resistance is calculated. Results (Fig.4 (right)) indeed have
a $\rho_{dc}$ discontinuity (above $\rho_{dc}$$\approx$7)
at the critical temperature $T_c$. 
At $T$$>$$T_c$, no cluster percolates leading to infinite
$\rho_{dc}$, but this can be easily fixed assuming a finite conductivity
among the many $Q$ clusters. Of course, it is not at all claimed
here that real manganites have Potts symmetry, this is just an 
illustration of concrete model Hamiltonians 
that presents the first-order transition described in the
previous section by cellular automata rules (rather than emerging from
a Hamiltonian approach).

\section{Conclusions}

In summary, a phenomenological
approach is proposed to include 
correlation and surface tension effects in percolation. 
The rounding of clusters induces
an abrupt first-order transition in the model,
consistent with results
of some experiments Fig.1(b).
The present calculation suggests that many manganites
do not belong to the standard percolation class, as believed by many
before, but correlation effects change the universality class inducing
a faster percolative process that may present hysteretic features
\cite{schiffer}. Note that the fairly ``random looking''
configuration Fig.2(g) is \emph{self-generated} \cite{selfgenerated} 
by our approach, its
phenomenological origin lies in phase competition, and 
is \emph{not} pinned by explicit disorder. 
Thus, the first-order percolative transition is predicted to 
appear particularly in nominally ``clean samples''
and for low intrinsic chemical disorder. If explicit disorder is added
to the system Fig.2(g), then the transition 
is expected to become continuous. More work is needed to fully
clarify the fascinating physics of CMR manganites.

\vskip 0.1cm

This work was supported in part by NSF grant DMR-0122523,
and also by MARTECH (FSU).


\begin{thebibliography}{99}


\bibitem{moreo} A. Moreo, S. Yunoki and E. Dagotto, Science {\bf 283},
2034 (1999).

\bibitem{uehara} M. Uehara, S. Mori, C. H. Chen and S.-W. Cheong, Nature
{\bf 399}, 560 (1999).

\bibitem{faeth} M. F\"ath, S. Freisem, A. A. Menovsky, Y. Tomioka,
J. Aarts and J. A. Mydosh, Science {\bf 285}, 1540 (1999).


\bibitem{review} E. Dagotto, T. Hotta and A. Moreo, Phys. Rep. {\bf 344}, 1
(2001). See also E. Dagotto, {\it Nanoscale Phase Separation and
Colossal Magnetoresistance}, Springer-Verlag, October 2002.

\bibitem{burgy} J. Burgy, M. Mayr, V. Martin-Mayor, A. Moreo and E. Dagotto,
Phys. Rev. Lett. {\bf 87}, 277202 (2001).

\bibitem{saitoh} E. Saitoh, Y. Tomioka, T. Kimura and Y. Tokura
J. Phys. Soc. Jpn. {\bf 69}, 2403 (2000).

\bibitem{baca} J. A. Fernandez-Baca, P. Dai, H. Kawano-Furukawa, H. Yoshizawa,
E. W. Plummer, S. Katano, Y. Tomioka, and Y. Tokura, Phys. Rev. B {\bf 66},
054434 (2002)

\bibitem{book} {\it Introduction to Percolation Theory}, by D. Stauffer
and A. Aharony, Taylor \& Francis publishers, 1994. 

\bibitem{khomskii} D. I. Khomskii and K. I. Kugel, cond-mat/0112340.

\bibitem{comment} The analog phenomenon --thin FM conductors turning
insulator in an insulating background-- also occurs.

\bibitem{bootstrap}
J. Chalupa, P. L. Leath and G. R. Reich, J. Phys. C {\bf 12}, L31 (1979).

\bibitem{chaves}
C. M. Chaves and B. Koiller, Physica A {\bf 218}, 271 (1995).
The Ising model is also related to percolation through the
Fortuin-Kasteleyn mapping \cite{fortuin}. See also
P. J. Bastiaansen and H. J. F. Knops, J. Phys. A {\bf 30}, 1791 (1997);
D. A. Wollman, M. A. Dubson and Q. Zhu, Phys. Rev. B {\bf 48}, 3713 (1993).

\bibitem{fortuin}
C. M. Fortuin and P. W. Kasteleyn, Physica {\bf 57}, 536 (1972).

\bibitem{stanley} H. A. Makse, S. Havlin, M. Schwartz and H. E. Stanley,
Phys. Rev. E {\bf 53}, 5445 (1996).

\bibitem{comment3} 
A. Weinrib, Phys. Rev. B {\bf 29}, 387 (1984) predicted
that correlations are relevant if $\gamma\nu < 2$, where $\nu$
is the correlation-length exponent for standard percolation.

\bibitem{vichniac} G. Y. Vichniac, Physica D {\bf 10}, 96 (1984).

\bibitem{cheese} B. I. Halperin, S. Feng and P. N. Sen, Phys. Rev. Lett.
{\bf 54}, 2391 (1985);
S. Feng, B. I. Halperin and P. N. Sen, Phys. Rev. B {\bf 35}, 197 (1987).

\bibitem{octavio} M. Octavio and C. J. Lobb, Phys. Rev. B {\bf 43},
8233 (1991).

\bibitem{comment2} However, some
configurations oscillate with period $2$. 

\bibitem{franklobb}
D. J. Frank and C. J. Lobb, Phys. Rev. B {\bf 37}, 302 (1988).

\bibitem{mayr} M. Mayr, A. Moreo, J. A. Verg\'es, J. Arispe, A. Feiguin
and E. Dagotto, Phys. Rev. Lett. 86, 135 (2001).

\bibitem{cluster} R. H. Swendsen and J. S. Wang, Phys. Rev. Lett.{\bf 58}, 
86 (1987); J. S. Wang and R. H. Swendsen, Physica A {\bf 167},
565 (1990).

\bibitem{baxter} R. J. Baxter, J. Phys. C, {\bf 6}, L445 (1973).

\bibitem{fortunato}
F. Peruggi, Physica A {\bf 141}, 140 (1987);
S. Fortunato and H. Satz, Nucl. Phys. B {\bf 623}, 493 (2002).

\bibitem{schiffer}
The complex dynamical behavior of manganites observed experimentally
(see, e.g., I. G. Deac, S. V. Diaz, B. G. Kim, S.-W. Cheong, and P. Schiffer
Phys. Rev. B {\bf 65}, 174426 (2002)) could have its
origin in the unusual percolative process described here.

\bibitem{selfgenerated} J. Schmalian and P. G. Wolynes, Phys. Rev. Lett.
{\bf 85}, 836 (2000). 


\end{thebibliography}
\end{document}